\documentclass[11pt,twoside]{article}
\usepackage{asp2010}
\usepackage{graphicx,wrapfig}

\resetcounters

\newcommand{\nue}{\ensuremath{\nu_{e}}}
\newcommand{\nuebar}{\ensuremath{\bar \nu_e}}
\newcommand{\numt}{\ensuremath{\nu_{\mu\tau}}}
\newcommand{\numtbar}{\ensuremath{\bar \nu_{\mu\tau}}}
\newcommand{\numu}{\ensuremath{\nu_{\mu}}}
\newcommand{\nutau}{\ensuremath{\nu_{\tau}}}
\newcommand{\numubar}{\ensuremath{\bar \nu_{\mu}}}
\newcommand{\nutaubar}{\ensuremath{\bar \nu_{\tau}}}
\newcommand{\mev}{\mbox{MeV}}
\newcommand{\isotope}[2]{$^{#2}$#1}
\newcommand{\gcc}{\ensuremath{{\mbox{g~cm}}^{-3}}}

\begin{document}

\title{Two- and Three-Dimensional Multi-Physics Simulations of Core Collapse Supernovae: A Brief Status Report and Summary of Results from the ``Oak Ridge'' Group}
\author{
Anthony Mezzacappa$^{1,2}$, Stephen W. Bruenn$^3$, Eric J. Lentz$^{1,2,4}$, W. Raphael Hix$^{1,4}$, O.E. Bronson Messer$^{4,5}$, J. Austin Harris$^{1}$, 
Eric J. Lingerfelt$^{6}$, Eirik Endeve$^{1,6}$, Konstantin N. Yakunin$^{2}$,  John M. Blondin$^{7}$, and Pedro Marronetti$^{8}$
\affil{
$^1$Department of Physics and Astronomy, University of Tennessee, Knoxville, TN 37996\\
$^2$Joint Institute for Computational Sciences, Oak Ridge National Laboratory, Oak Ridge, TN, 37831\\
$^3$Department of Physics, Florida Atlantic University, Boca Raton, FL 33431\\
$^4$Physics Division, Oak Ridge National Laboratory, Oak Ridge, TN 37831\\
$^5$National Center for Computational Sciences, Oak Ridge National Laboratory, Oak Ridge, TN 37831\\
$^6$Computer Science and Mathematics Division, Oak Ridge National Laboratory, Oak Ridge, TN 37831\\
$^7$Department of Physics, North Carolina State University, Raleigh, NC 27695\\
$^8$Physics Division, National Science Foundation, Arlington, VA 22230
}
}

\begin{abstract}
We summarize the results of core collapse supernova theory from one-, two-, and three-dimensional models and 
provide a snapshot of the field at this time. We also present results from the ``Oak Ridge'' group in this context.
Studies in both one and two spatial dimensions define the {\it necessary} physics that must be included in core 
collapse supernova models: a general relativistic treatment of gravity (at least an approximate one), spectral 
neutrino transport, including relativistic effects such as gravitational redshift, and a complete set of neutrino weak 
interactions that includes state-of-the-art electron capture on nuclei and energy-exchanging scattering on electrons and nucleons. 
Whether or not the necessarily approximate treatment of this physics in current models that include it is {\it sufficient} 
remains to be determined in the context of future models that remove the approximations. We summarize the results 
of the Oak Ridge group's two-dimensional supernova models. In particular, we demonstrate that robust neutrino-driven 
explosions can be obtained. We also demonstrate that our predictions of the explosion energies and remnant neutron 
star masses are in agreement with observations, although a much larger number of models must be developed
before more confident conclusions can be made. We provide  preliminary results 
from our ongoing three-dimensional model with the same physics. 
Finally, we speculate on future outcomes and directions.
\end{abstract}

\section{Introduction}

We have come a long way since the first proposal that core collapse supernovae could be driven by the emission of neutrinos from the post-bounce stellar core \citep{cowh66}. Progress has been marked by (i) significant improvements in the treatment of the neutrino transport [e.g., see \cite{limeme04}; see also \cite{otbude08} and \cite{suya12}], (ii) an expansion of the list of neutrino weak interactions included in the transport and an increase in the sophistication of the models used to describe such interactions \citep{reprla98,busa98,hara98,lamasa03,bujake03}, (iii) the use of an increasing number of equations of state \citep{lasw91,shtooy98,shhooc11,HeFiSc12,fusuya13,mepoma13}, (iv) the inclusion of general relativity \citep{limeme04,mujama12,brmehi13,kukota12,otabmo13}, and, last but certainly not least, (v) the migration of multiphysics core collapse supernova models from one spatial dimension with spherical symmetry imposed [e.g., see \cite{lememe12}], to two spatial dimensions with axisymmetry imposed [e.g., see \cite{hebehi94,sukota10,mujahe12,mujama12,brmehi13,dobuzh14}], to three spatial dimensions with no imposed symmetries [e.g., see \cite{frwa04,fryo07,kukota12,takosu13,hamuwo13}]. Recent progress, especially in the context of two-dimensional axisymmetric models, is promising \citep{sukota10,mujama12,brmehi13}. The combination of neutrino heating, neutrino-driven convection, and the standing accretion shock instability (SASI) has led to first-principles explosions across a range of stellar progenitors.

\section{Lessons from Spherical Symmetry}

\begin{wrapfigure}[18]{r}{3.00in}
\includegraphics[width=3.00in]{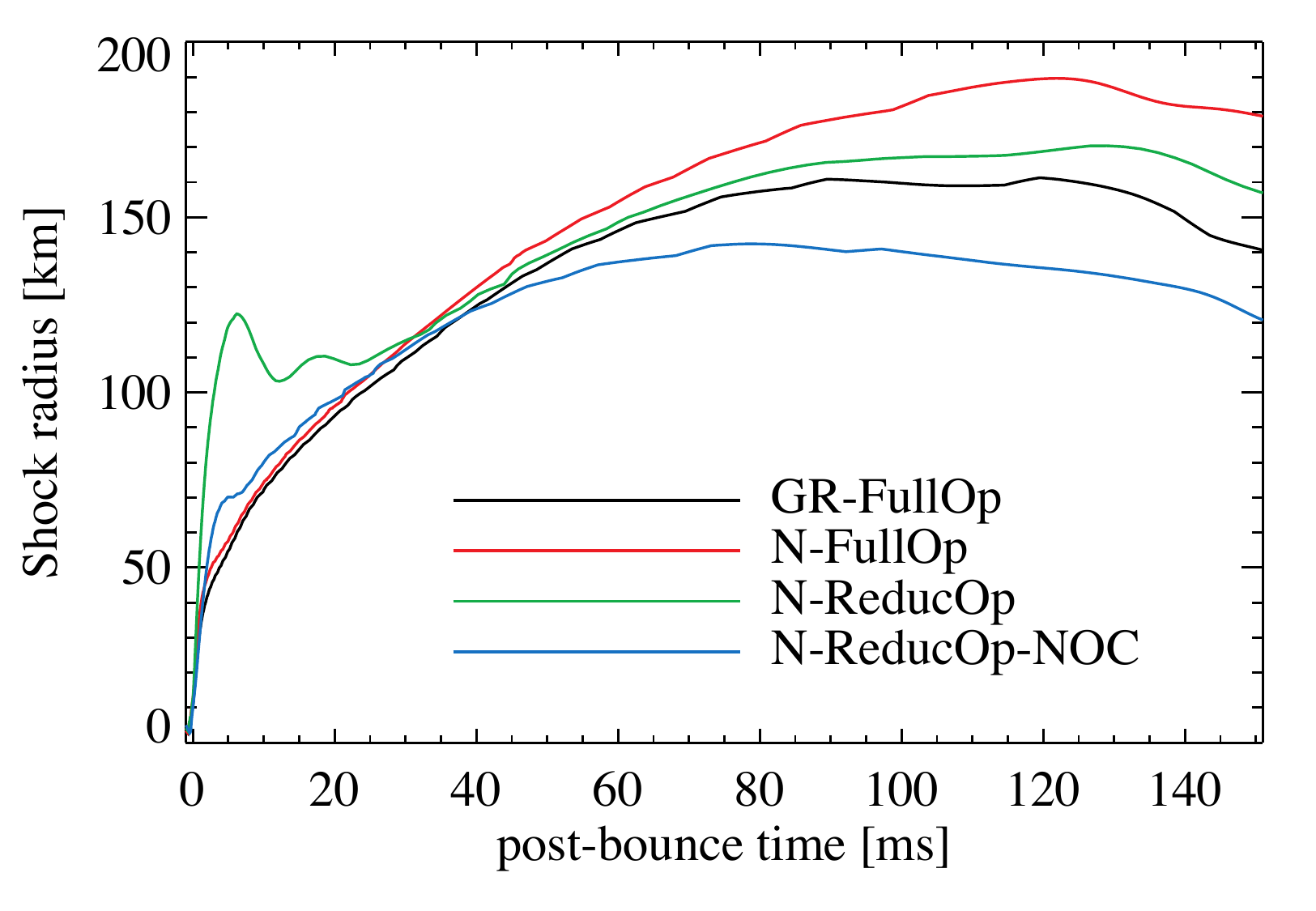}
\caption{Shock trajectories in km, versus time after bounce, for models with decreasing physics.
\label{fig:shock}
}
\end{wrapfigure}

Recent studies carried out in the context of general relativistic, spherically symmetric core collapse supernova models with Boltzmann neutrino transport demonstrate that (i) a general relativistic treatment of gravity, (ii) special and general relativistic corrections to the neutrino transport, such as the gravitational redshift of neutrinos, and (iii) the use of a complete set weak interactions and a realistic treatment of those interactions are indispensable \citep{lememe12}. As shown in Figure 1, the impact of moving to a Newtonian description of gravity from a fully general relativistic treatment has a significant impact on the shock trajectory. The Newtonian simulation neglects general relativity in the description of gravity {\it per se}, as well as general relativistic transport effects such as gravitational redshift. Thus, the switch from a general relativistic description to a Newtonian description impacts more than just the treatment of gravity. In turn, if we continue to simplify the model, this time reducing the set of weak interactions included and the realism with which these weak interactions are included, we see a further significant change in the shock trajectory, with fundamentally different behavior early on after bounce. In this instance, we have neglected the impact of nucleon correlations in the computation of electron capture on nuclei, energy exchange in the scattering of neutrinos on electrons, corrections due to degeneracy and nucleon recoil in the scattering of neutrinos on nucleons, and nucleon--nucleon bremsstrahlung. Finally, if we continue to simplify the neutrino transport by neglecting special relativistic corrections to the transport, such as the Doppler shift, we obtain yet another significant change. The spread in the shock radii at $t>$120 ms after bounce is approximately 60 km. Its relative fraction of the average of the shock radii across the four cases at $t>$ 120 ms is $>$33\%. The conclusions of these studies are corroborated by similar studies carried out in the context of two-dimensional multi-physics models \citep{mujama12}. In essence, these studies establish the {\it necessary} physics that must be included in core collapse supernova models. Whether or not the current treatments of this physics in the context of two- and three-dimensional models is {\it sufficient} remains to be determined.

\section{The CHIMERA Code}

CHIMERA is a parallel, multi-physics code built specifically for multidimensional simulation of CCSNe.
It is the chimeric combination of separate codes for hydrodynamics and gravity; neutrino transport and opacities; and a nuclear EoS and reaction network, coupled by a layer that oversees data management, parallelism, I/O, and control.
The hydrodynamics are modeled using a dimensionally-split, Lagrangian-Remap (PPMLR) scheme  \citep{CoWo84} as implemented in VH1 \citep{HaBlLi12}.
Self-gravity is computed by multipole expansion \citep{MuSt95}. In the GR case, the Newtonian monopole is replaced with a GR monopole  \citep[][Case~A]{MaDiJa06}.
Neutrino transport is computed in the ``ray-by-ray-plus'' (RbR+) approximation \citep{buraja03}, where an independent, spherically symmetric transport solve is computed for each ``ray'' (radial array of zones with the same $\theta$, $\phi$).
Neutrinos are advected laterally (in the $\theta$ and $\phi$ directions) with the fluid and contribute to the lateral pressure gradient where $\rho>10^{12}\,\gcc$.
The transport solver is an improved and updated version of the multi-group flux-limited diffusion transport solver of \citet{Brue85} enhanced for GR \citep{BrDeMe01}, with an additional geometric flux limiter to prevent the over-rapid transition to free streaming of the standard flux-limiter.  All $O(v/c)$ observer corrections have been included.
CHIMERA solves for all three flavors of neutrinos and antineutrinos with four coupled species: \nue, \nuebar, $\numt=\{\numu,\nutau\}$, $\numtbar=\{\numubar,\nutaubar\}$, with typically 20 energy groups each for $\alpha\epsilon =  4-250~\mev$, where $\alpha$ is the lapse function and $\epsilon$ is the comoving-frame group center neutrino energy.
Our standard, modernized, neutrino--matter interactions include emission, absorption, and non-isoenergetic scattering on free nucleons \citep{reprla98}, with weak magnetism corrections \citep{Horo02}; emission/absorption (electron capture) on nuclei \citep{lamasa03}; isoenergetic scattering on nuclei, including ion-ion correlations; non-isoenergetic scattering on electrons and positrons; and pair emission from $e^+e^-$-annihilation \citep{Brue85} and nucleon-nucleon bremsstrahlung \citep{hara98}.
We also have available as alternatives simpler versions of absorption, scattering, and isoenergetic scattering on neutron and protons and electron capture on nuclei as specified in \citet{Brue85}.
CHIMERA generally utilizes the $K = 220$~\mev\ incompressibility version of the \citet{lasw91} EoS for  $\rho>10^{11}\,\gcc$ and a modified version of the \citet{Coop85} EoS for  $\rho<10^{11}\,\gcc$, where nuclear statistical equilibrium (NSE) applies.
Most CHIMERA simulations have used a 14-species $\alpha$-network ($\alpha$, \isotope{C}{12}-\isotope{Zn}{60}) for the non-NSE regions \citep[XNet][]{HiTh99a}.
To aid the transition between the network and NSE regimes, we have constructed a 17-species NSE solver to be used in place of the Cooperstein EoS for electron fractions $Y_{\rm e}> 0.46$.
An extended version of the Cooperstein electron--photon EoS is used throughout.

\begin{wrapfigure}[42]{r}{3.00in}
\includegraphics[width=3.00in]{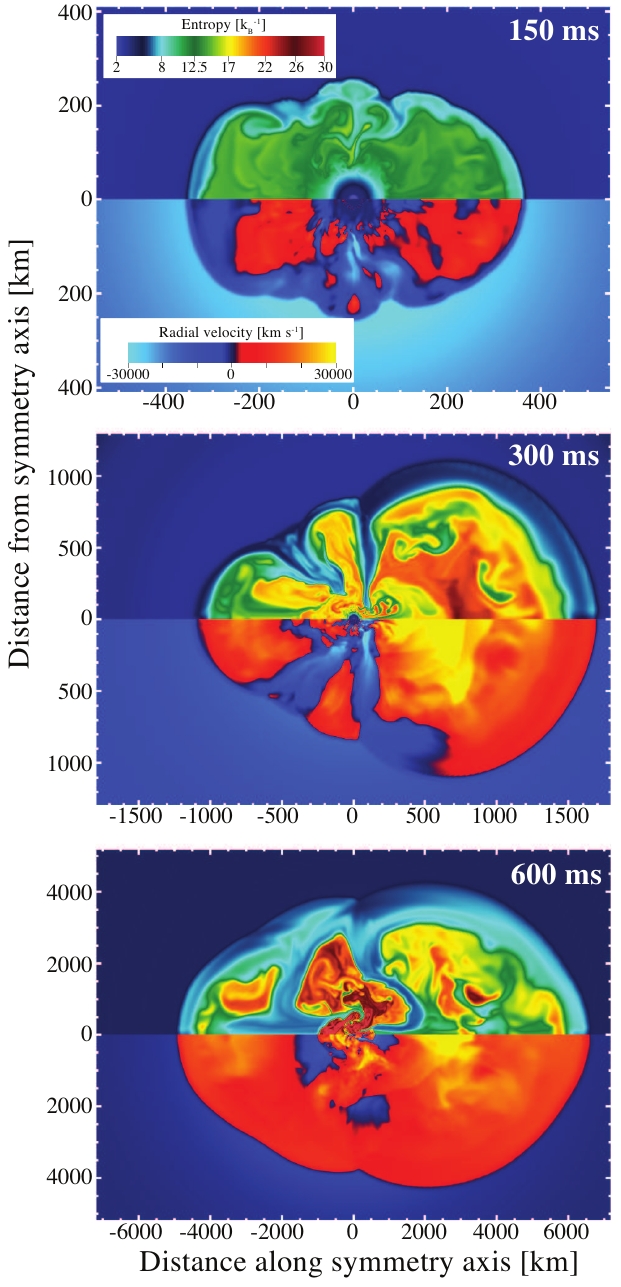}
\caption{
Evolution of the entropy (upper half) and radial velocity (lower half) st 150, 300, and 600~ms after bounce  for B12-WH07.  The  scale grows in time to capture the expansion of the supernova shockwave, but the colormaps remain constant. 
\label{fig:entropy}
}
\end{wrapfigure}

During evolution the radial zones are gradually and automatically repositioned during the remap step to track changes in the radial structure.
To minimize restrictions on the time step from the Courant limit, the lateral hydrodynamics for a few inner zones are ``frozen'' during collapse, and after prompt convection fades the laterally frozen region expands to the inner 8--10~km.
In the ``frozen'' region, lateral velocities are set to 0 and the lateral hydrodynamic sweep is skipped. The full radial hydrodynamics and neutrino transport are always computed to the center of the simulation for all rays.

The supernova code most closely resembling CHIMERA is the VERTEX code developed by the MPA group \citep{buraja03,buraja06,bujara06,MaJa09}. This code utilizes a ray-by-ray-plus approach to neutrino transport, solving the first two multifrequency angular moments of the transport equations with variable Eddington factors, which are in turn obtained by solving, at intervals, a 1D model Boltzmann equation along each radial ray.

\section{Status of State-of-the-Art Two- and Three-Dimensional Models}

First-principles (nonparameterized) neutrino-driven explosions have been obtained by several groups for a range of initial progenitor masses 
\citep{sukota10,mujama12,brmehi13}. 
The outcomes of these simulations fuel continued belief that core collapse supernovae can be neutrino driven. Two of these sets of simulations \citep{mujama12,brmehi13,Bretal14} include RbR spectral neutrino transport with relativistic corrections (e.g., gravitational redshift), a complete set of neutrino weak interactions, and approximate general relativity -- i.e., {\it necessary} realism. They agree qualitatively, although there are important quantitative differences that need to be resolved. In particular, the explosion energies obtained by the Max Planck group are significantly lower than those obtained by the Oak Ridge group. Figure 2 shows the evolution of two quantities, entropy in the upper panel and radial velocity in the lower panel, for our two-dimensional model initiated from a 12 M$_{\odot}$ progenitor, at several stages after stellar core bounce. The change in the radial scale as we move downward in the figure is clear indication of explosion. The prolate shock shape is in part the result of the dominant $l=1$, sloshing SASI mode in this case, which is in part the result of the imposition of axisymmetry (other nonlinear SASI modes -- e.g., the $m=1$, spiral mode -- require three spatial dimensions). It is also apparent at late times in the sequence the explosion becomes self similar. This is consistent with the findings of \cite{FeMuFo13}. Figure 3 shows the evolution of the explosion energies in our four two-dimensional models, beginning with 12, 15, 20, and 25 M$_\odot$ progenitors. Several things are evident from the plot: (1) Robust neutrino-driven explosions with energies $O(1 \rm{B})$ are possible (1 B $= 1\times 10^{51}$ erg). (2) With the exception of the 12 M$_\odot$ case, the explosion energy continues to grow well past 1 s post stellar-core-bounce. (3) The explosion energy in the 12 M$_\odot$ case has leveled off at a value $\sim 0.35$ B, given the falloff in density in the progenitor and, consequently, the falloff in the mass accretion onto the proto-neutron star, which powers the neutrino luminosities. Figure 2 clearly shows the massive accretion funnels, most evident at $\sim 300$ ms after bounce (along the ``equator'' in the plane orthogonal to the axis of symmetry) impinging on the proto-neutron star surface. The SASI is responsible for the creation and structure of these accretion funnels, which sustain significant mass accretion onto the proto-neutron star long after stellar core bounce. Significant accretion is still evident at $\sim 600 $ ms after bounce, and continues beyond this in all but the 12 M$_\odot$ case. In Figure 5 we plot observed explosion energies for a number of core collapse supernovae, along with our predicted explosion energies at the end of each simulation. The red arrows indicate that our explosion energies are still growing when the runs are stopped, although the growth in the 12 M$_\odot$ case is negligible. The growth in the explosion energy in both the 20 and 25 M$_\odot$ models is still significant when we stopped these runs, indicating that our simulations must be carried out further in time before we can make accurate predictions of the final energies generated. Our predicted energy for our 15 M$_\odot$ model is comparable to observed energies for progenitors of approximately the same mass. Unfortunately, observational data above 18 M$_\odot$ and between 9 and 12 M$_\odot$ are scant. The growth of the proto-neutron star baryonic masses as a result of this continued accretion onto it is shown in Figure 5. Our final proto-neutron star masses range between 1.5 and 1.9 M$_\odot$. While continued accretion continues to fuel appreciably the neutrino luminosities at late times in all but the lightest progenitor, and hence the explosion energy, it does not have appreciable impact on the proto-neutron star mass.

Few three-dimensional multiphysics models with necessary realism (as defined above) have been performed. Notable among these are the recently published models of \cite{takosu12} and \cite{hamuwo13}, although the Takiwaki et al. models are Newtonian. Preliminary results from the Oak Ridge group in the context of models similar to the Garching group -- i.e., including general relativity -- are presented here. Figure 6 is a snapshot of a two-dimensional slice of our ongoing three-dimensional model at 267 ms after bounce \citep{Leetal14}. Shown is the stellar core entropy. The shock wave is clearly outlined by the jump in entropy across it. Neutrino-driven convection is evident in the slice. Hotter (red) rising plumes bring neutrino-heated material up to the shock, while cooler (yellow) down flows replace the fluid below. Distortion of the shock away from axisymmetry and the nonaxisymmetric patterns of convection beneath the shock are also evident. Conclusive evidence for $l=1$, ``sloshing'' and $m=1$, ``spiral'' modes of the SASI will require a modal analysis, although the two-dimensional slice clearly does not rule out either mode. 

\section{The 800 Pound Gorilla: Neutrino Transport}

One avenue through the history of core collapse supernova theory is the avenue we just took: through the development of sophisticated one-, two-, and three-dimensional multiphysics models. But it is important to deconvolve the improvements of multiphysics models from the improvements of individual components, especially the neutrino transport. At present, most modeling efforts still circle around the idea that core collapse supernovae are neutrino driven. From this standpoint alone, the neutrino sector in core collapse supernova models is arguably the most important. Adding to this, the demonstrated sensitivity of the predicted outcomes of supernova models to the treatment of the neutrino physics \citep{lememe12,mujama12} argues for a treatment with necessary, if not sufficient, realism.

\begin{wrapfigure}[19]{r}{3.00in}
\includegraphics[width=3.00in]{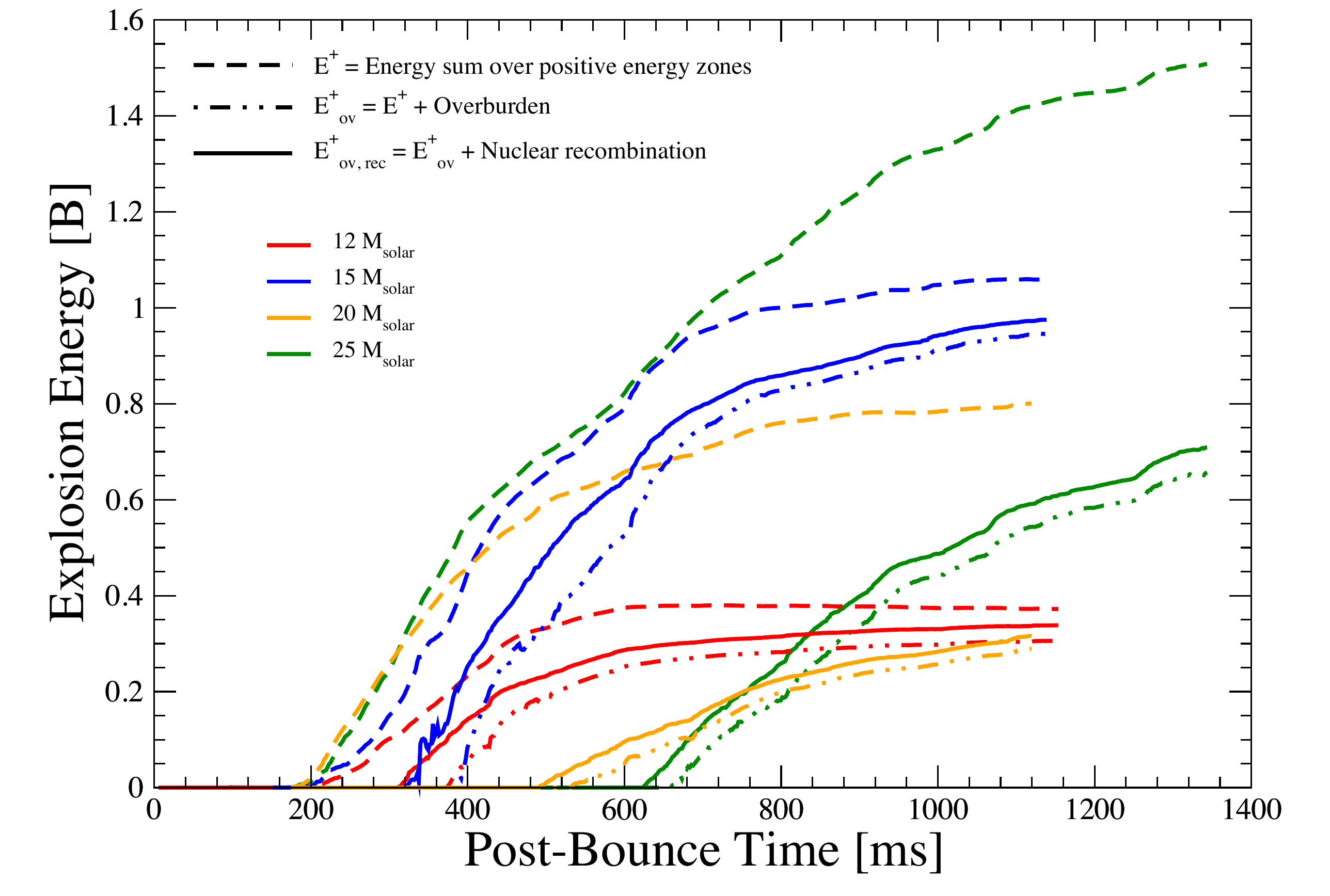}
\caption{
Time evolution of the explosion energies in each of our 4 two-dimensional models, beginning with 12, 15, 20, and 25 M$_\odot$ progenitors.
\label{fig:3Dentropy}
}
\end{wrapfigure}

Understandably, treatments of neutrino transport have advanced most in the context of spherically symmetric models of core collapse supernovae \citep{limeme04,liraja05}. In this context, a complete general relativistic description of Boltzmann kinetics has been deployed, with state-of-the-art weak interaction physics \citep{limeme04}. This remains the bellwether for what ultimately must be achieved in three spatial dimensions. Boltzmann kinetics is evolved in phase space. Hence, a spatially one-dimensional problem becomes a three-dimensional phase-space problem (space, one direction cosine, and energy). In three spatial dimensions, the dimensionality explodes to six phase-space dimensions (space, two direction cosines, and energy). While we are fortunate to have achieved this milestone at least in the context of spherical symmetry, the implementation of six-dimensional Boltzmann kinetics must await supercomputer platforms that will be deployed a decade from now, and perhaps beyond if sufficient spatial resolution is required. [Notable work towards this goal has been completed by \cite{otbude08} and \cite{suya12}.] Even then, fundamental memory limitations we might expect in future supercomputer architectures may render a Boltzmann kinetics treatment difficult, given the memory footprint of such a treatment. In light of these inescapable limitations, how should we proceed?

Two practical approaches present themselves: 

\begin{wrapfigure}[23]{r}{3.00in}
\includegraphics[width=3.00in]{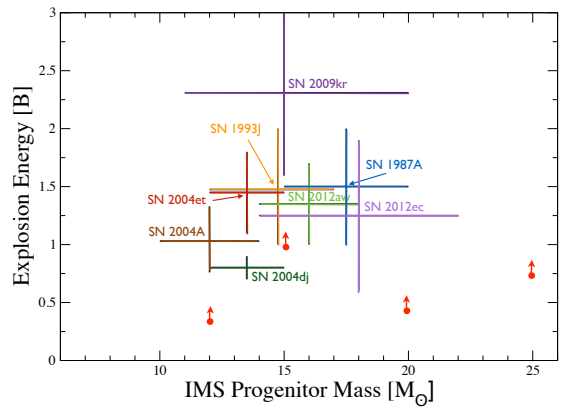}
\caption{
Observed explosion energies for a number of core collapse supernovae, along with predicted explosion energies from our 12, 15, 20, and 25 M$_\odot$ progenitor models (red dots). The arrows indicate our energies are still increasing at the end of each run.
\label{fig:3Dentropy}
}
\end{wrapfigure}

(1) The use of the Ray-by-Ray (RbR) approximation \citep{buraja03}. In this approximation, sophisticated neutrino transport codes that have been developed and deployed in the context of spherically symmetric models are used in spatially two- and three-dimensional models. The results we present here were obtained using this approach. In essence, in the RbR approximation we solve N spherically symmetric problems, where N is the number of angular spatial bins in our two- and three-dimensional models. As shown in Figure 7, the base of each ray on the neutrinosphere marks the surface midpoint of a partial surface subtended by the backward causal cone of an arbitrary point above the neutrinosphere. In particular, the neutrino heating and cooling at that arbitrary point is computed using the neutrino distributions obtained by assuming spherical symmetry for the portion of the surface subtended by the cone. N such problems are solved, one for each ray emanating from the neutrinosphere. The fundamental limitation of the RbR approach can easily be seen: If the point at which the radial ray intersects the neutrinosphere is at a higher temperature (e.g., from the impact of an accretion funnel), the RbR approximation will treat the entire subtended surface as being at that temperature, potentially overestimating the neutrino heating at point 1 above the proto- neutron star surface. Similarly, if the radial-ray base point is at a lower temperature (despite the fact there may be a hot spot within the subtended portion of the neutrinosphere surface), the RbR approximation treats the whole subtended surface as cold, thereby potentially underestimating the neutrino heating at point 2 above the surface. In essence, the RbR approach exaggerates angular variations. However, as a function of time, if hotter material accreted onto the proto-neutrino star surface spreads over the surface quickly and if the accretion funnels feeding such accretion onto this surface themselves move in spatial angle, this limitation is somewhat ameliorated. A precise analysis of the shortcomings of the RbR approach will require direct comparison with a non-RbR treatment [e.g., see \cite{sutama14,dobuzh14}], hopefully of equal sophistication in terms of the physics included (general relativity, weak interactions, etc.). In the meantime, the RbR is a reasonable approach to take, particularly as it enables the inclusion of other necessary realism in the models.

(2) The use of a neutrino-direction-cosine--integrated (moment) approach [e.g., see \cite{dobuzh14}]. In this case, we preserve the neutrino energy dimension -- i.e., we preserve spectral neutrino transport -- but we reduce the dimensionality of the problem by integrating out the neutrino angular degrees of freedom and solve for the lowest one or two spectral angular moments of the neutrino distribution. This approach holds the greatest promise to advance core collapse supernova theory over the next decade. Such an approach is practical in terms of both floating point operations required to solve the equations of general relativistic spectral neutrino radiation hydrodynamics and the memory footprint required to host such systems on today's architectures. 

\section{Neutrino-Driven Convection versus the SASI}

\begin{wrapfigure}[19]{r}{3.00in}
\includegraphics[width=3.00in]{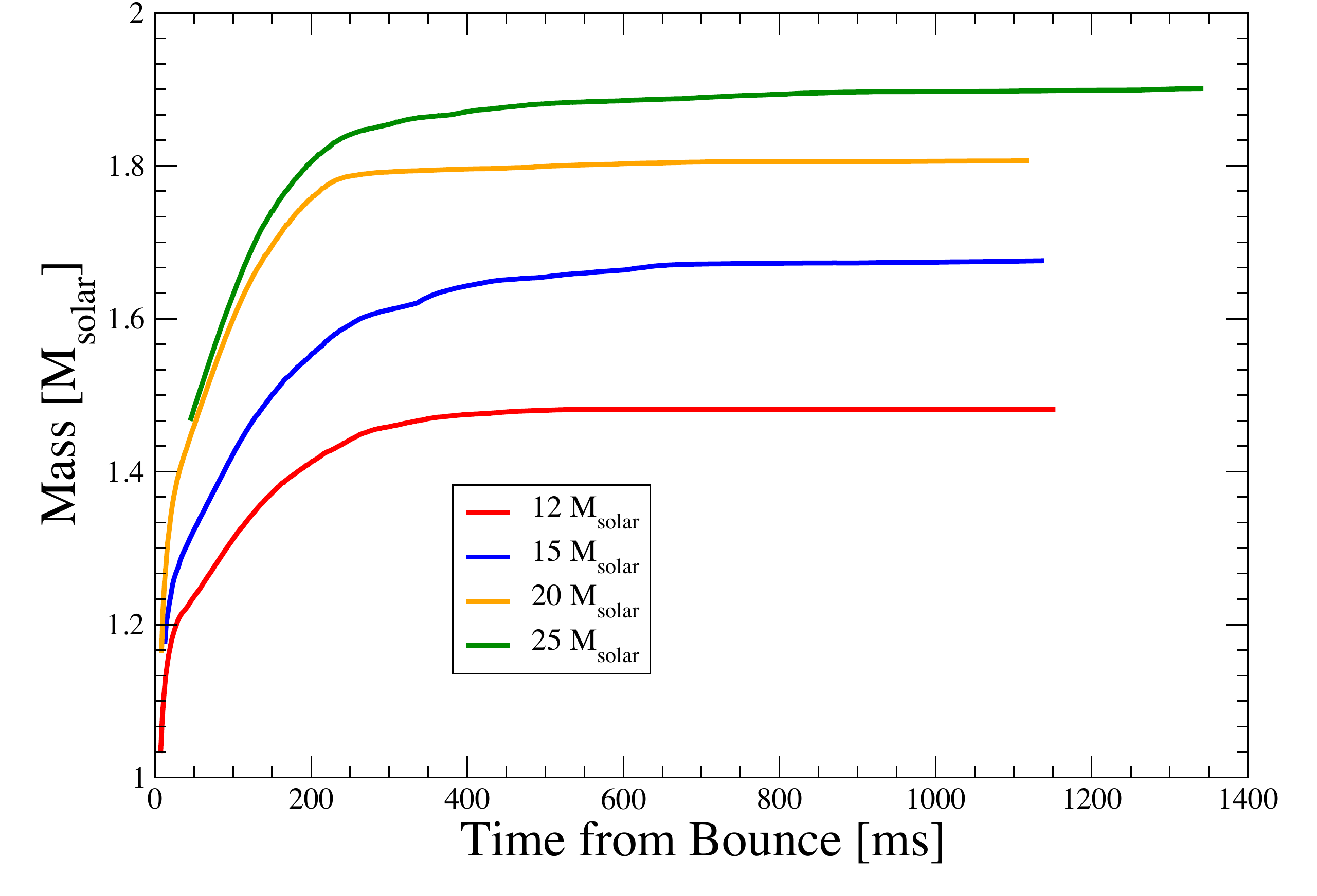}
\caption{
Time evolution of the proto-neutron star (baryonic) mass in each of our 4 two-dimensional models, beginning with 12, 15, 20, and 25 M$_\odot$ progenitors.
\label{fig:3Dentropy}
}
\end{wrapfigure}

The introduction of a second spatial dimension in core collapse supernova models \citep{hebehi94} freed models from the fundamental limitation of spherically symmetric models: the inability to support mass accretion, which powers the neutrino luminosities, while simultaneously supporting explosions. Once an explosion is initiated in the context of spherical symmetry, mass accretion is cut off. Spherical symmetry does not allow simultaneous accretion down flows and explosive outflows. The move to axisymmetry, in turn, removed this fundamental limitation and enabled this for the first time. Key to early two-dimensional models was the appearance of neutrino-driven convection directly below the supernova shock wave \citep{hebehi94,buhafr95,jamu96}, which aids neutrino heating in a number of ways. However, early two-dimensional simulations were not performed sufficiently long to enable the standing accretion shock instability (SASI) to develop fully. This is an instability of the shock wave itself and results in a ``sloshing'' mode in two spatial dimensions \citep{blmede03} and in both sloshing and spiral modes in three dimensions \citep{blme07}. The development timescales for neutrino-driven convection and the SASI, and in particular how they compare, depend on the hydrodynamic conditions in the stellar core and are therefore a function of the progenitor mass. Recently, multiple groups have endeavored to understand the relative roles of these two important instabilities in the core collapse supernova explosion mechanism \citep{nobual10,hamamu12,dobumu13,mudobu13,hamuwo13}. Multi-physics simulations conducted by the Oak Ridge and Max Planck groups suggest that neutrino-driven convection develops first in the context of lower progenitor mass (e.g., 11--12 M$_\odot$), simultaneously for intermediate progenitor masses (e.g., 15 M$_\odot$), and after the SASI in more massive progenitors (e.g., 25--30 M$_\odot$) \citep{mujahe12,mujama12,brmehi13}. Moreover, the SASI seems to be particularly important for the more massive progenitors. In short, as has been suggested by a number of authors and corroborated in the more detailed studies conducted by the Garching and Oak Ridge groups, both neutrino-driven convection and the SASI are present in post-bounce stellar cores and each plays a role in neutrino shock revival. Nonetheless, the extension of current two- and three-dimensional models to significantly longer post bounce times, and as a result, the entry of the SASI in the supernova dynamics, was unquestionably the next major step forward after the introduction of additional spatial dimensions in the first two-dimensional core collapse supernova simulations performed two decades ago.

\section{The Future}

\begin{wrapfigure}[23]{r}{3.00in}
\includegraphics[width=3.00in]{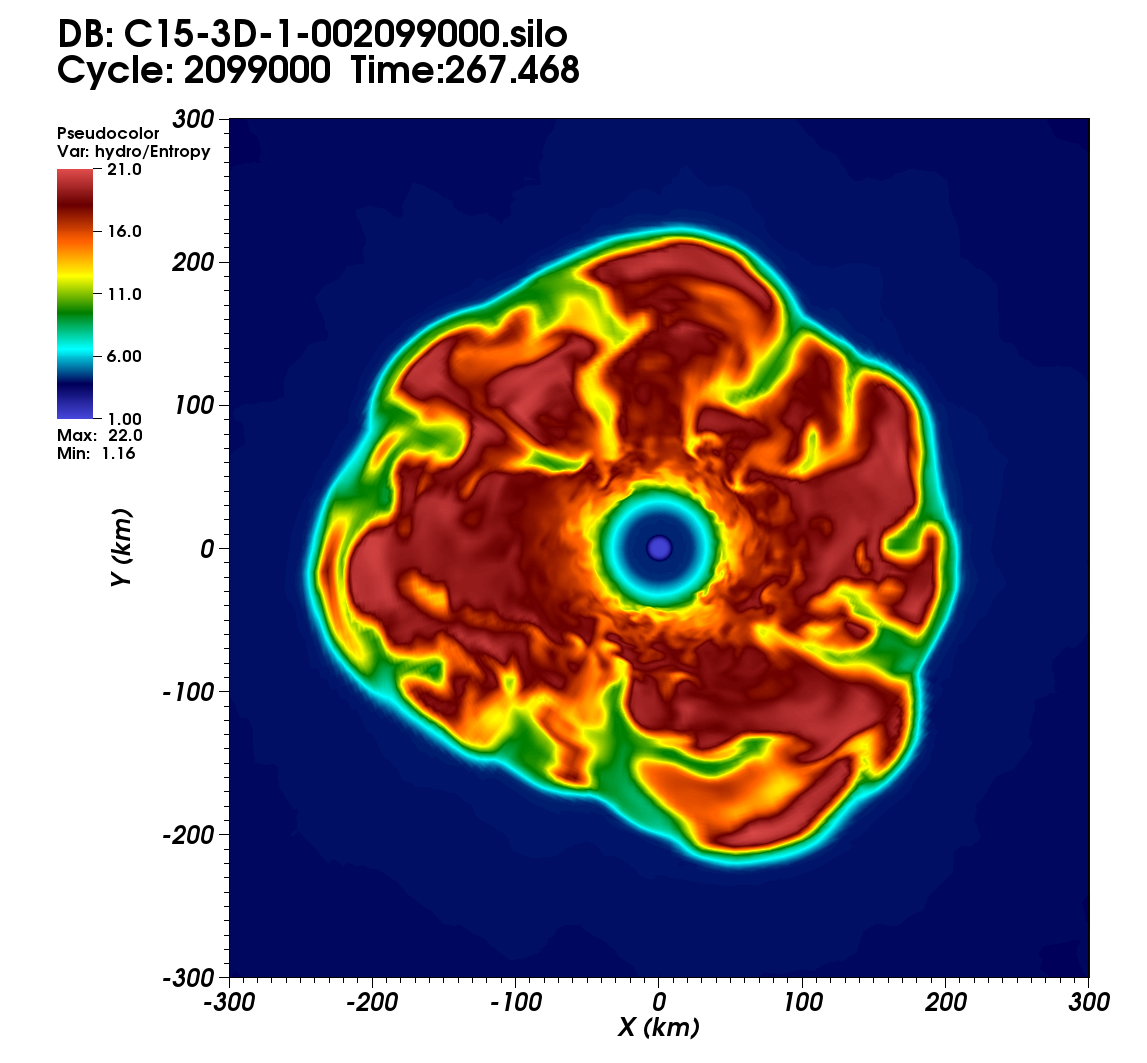}
\caption{
Snapshot of the equatorial cross section of the entropy in our ongoing three-dimensional simulation for the 15 M$_\odot$ case at 267 ms after bounce.
\label{fig:3Dentropy}
}
\end{wrapfigure}

Of course, significant work remains. First, many more two-dimensional models must be produced so (1) our understanding of core collapse supernova dynamics in the context of two-dimensional models becomes as well developed as our knowledge of this dynamics in the context of one-dimensional models and (2) our two-dimensional models can continue to provide guidance, and provide more complete guidance, to ongoing three-dimensional modeling efforts. There is a practical benefit here as well. The cost of current state-of-the-art three-dimensional multi-physics models (with {\it necessary} realism) will not permit very many models to be developed. Therefore, our best hope of spanning progenitor mass and characteristics, sensitivity to input microphysics, etc. will for some time rely on our ability to develop a compendium of two-dimensional supernova models. Nonetheless, the few three-dimensional models the community will bring forward at this time will necessarily be the arbiters of our ability to model and, in turn, understand core collapse supernovae. The first key question that will be addressed as these models come to fruition is: Are robust neutrino-driven explosions obtained in the context of three-dimensional core collapse supernova models with necessary (if not sufficient) realism? If the answer is no, we may be forced to conclude that physics is still missing from the models. It is possible, though perhaps not likely, removing current approximations in the models (e.g., the use of RbR neutrino transport), and making other improvements (e.g., increasing the spatial resolution), may fundamentally alter the outcomes. It is more likely we are missing something essential, the most obvious being to initiate our three-dimensional simulations with three-dimensional progenitors [e.g., see \cite{armevi14}]. Recent work indicates that precollapse structures that will likely be present in three-dimensional late-stage stellar evolution models may have significant impact on stellar core collapse and postbounce dynamics \citep{coot13}. 

\begin{wrapfigure}[39]{r}{2.50in}
\includegraphics[width=3.00in]{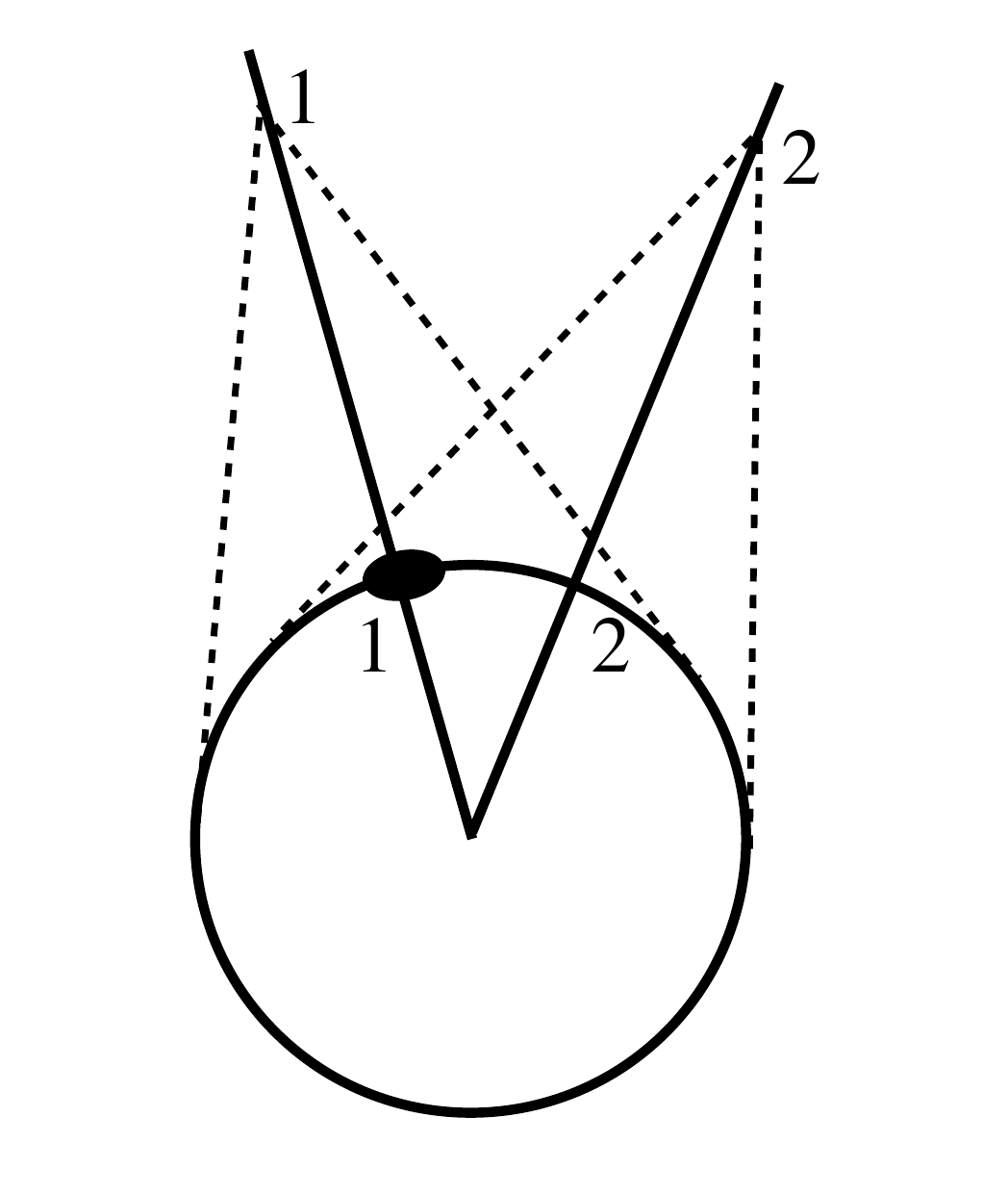}
\caption{A depiction of the Ray-by-Ray (RbR) approach. Each ray corresponds to a separate spherically symmetric problem. In the limit of spherical symmetry, the RbR approach is exact. Each ray solve gives what would be obtained in a spherically symmetric solve for conditions at the base of the ray, on the proto-neutron star surface. For a {\it persistent} hot spot, such as the one depicted here, the RbR approximation would overestimate the angular variations in the neutrino heating at the points 1 and 2 above the surface. 
\label{fig:rbi}
}
\end{wrapfigure}

Unless significant rotation is present, magnetic fields are not likely to be enhanced to dynamically dominant levels [e.g., see \cite{saya14} and references cited therein]. Recent observations of $^{44}$Ti in CassiopeiaA \citep{grhabo14} rule out a highly asymmetric bipolar explosion that would result from a rapidly rotating progenitor. 

If we are to continue to pursue a neutrino-driven mechanism, it may be logical to assume we are still missing something in the neutrino sector. Motivated by the experimental and observational measurement of neutrino mass, recent efforts to explore its impact on neutrino transport in stellar cores have uncovered new and increasingly complex physical scenarios \citep{dufuqi10,chcafr12,chcafr13,vlfuci13}.

On the other hand, if current three-dimensional models continue to show evidence of neutrino-driven explosions, a different question will arise: What relative role do neutrino-driven convection and the SASI play? This question naturally arises given the evidence that neutrino-driven convection and the SASI contribute in different (relative) ways as a function of progenitor mass \citep{mujama12,brmehi13}, but is particularly pressing in light of evidence that the energy in long-wavelength modes of the SASI may be sapped by the very turbulence the SASI seeds as a result of the significant shear between counterrotating flows induced by its $m=1$ spiral mode in three dimensions \citep{EnCaBu12}.

These are exciting times. The recent progress made, the growing capability offered by present-day supercomputers, and the extant capabilities in our ability to observe a Galactic core collapse supernova in both neutrinos and gravitational waves, in addition to our ability to observe such a supernova in detail across the electromagnetic spectrum, encourage us that a resolution of this long-standing astrophysics problem is possible and, in light of the scientific payload that will be provided by a Galactic event, imperative.

\section{Acknowledgements}

We acknowledge support from the DOE Office of Nuclear Physics and Office of Advanced Scientific Computing Research, NASA's Astrophysics Theory Program (grant no. NNH11AQ72I), and the NSF PetaApps Program (grant no. OCI-0749242). We also acknowledge compute resources on Jaguar and Titan at the Oak Ridge Leadership Computing Facility and on Hopper at the National Energy Research Scientific Computing Center, through DOE's Innovative and Novel Computational Impact on Theory and Experiment (INCITE) Program, and on Kraken, at the National Institute for Computational Sciences, through NSF's XRAC Program. The CHIMERA team would like to dedicate this paper to Stirling Colgate, whose passing occurred during its preparation.
 

\end{document}